\documentclass[prl,aps,twocolumn,floats,showpacspsfig]{revtex4} 
\usepackage{epsfig}

\def\be{\begin{equation}}
\def\ee{\end{equation}}
\def\bdm{\begin{displaymath}}
\def\edm{\end{displaymath}}
\def\bea{\begin{eqnarray}}
\def\eea{\end{eqnarray}}

\begin{document}
\draft
\title{Theory of Optical Transmission Through a Near-Field Probe With a 
Dissipative Matter in its Core}
\author{V. S. Lebedev, T. I. Kuznetsova  and A. M. Tsvelik $^*$}
\affiliation{P.N.~Lebedev Physical Institute, Leninsky prospect 53,  
Moscow 119991, Russia\\
$^*$  Department of  Physics, Brookhaven 
National Laboratory, Upton, NY 11973-5000, USA}
\date{\today}

\begin{abstract}
We develop a theory of light transmission through an  aperture-type 
near-field optical probe with a dissipative matter in its semiconducting core 
described by  a complex frequency-dependent dielectric function. 
We evaluate the near-field transmission coefficient of a  metallized 
silicon probe with a large taper angle of  
in the visible and near-infrared wavelength range. It is shown that in this 
spectral range  the use of a short silicon probe  instead of a glass one 
allows to achieve a  strong (up to 10$^2-10^{3}$) enhancement in the 
transmission efficiency.

\end{abstract}
\maketitle

The development of scanning near-field optical microscopy (SNOM) has 
already led to enormous progress in studies of different nanoscale 
phenomena. Among other things, this technique is widely used for image 
formation of various nanoobjects, single molecule fluorescence detection,  
and for laser-induced ablation of a sample close to the tip apex 
(see \cite{Hecht2000, Girard2000}). 
In the recent years SNOM methods have been  employed in studies of 
quantum dots \cite{Kawazae2002} and single-walled carbon nanotubes 
\cite{Hartschuh2003}. The near-field optical microscopy has opened a 
possibility for direct observation of fine features in the self-focusing 
effect \cite{Song2000}, for imaging the light propagation in photonic 
crystal waveguides \cite{Bozhevolnyi2002} and the electromagnetic local 
density of states of optical corrals \cite{Chicanne2002}.

To improve efficiency of the aperture probes 
for the SNOM 
technique one needs simultaneously to increase their transmittance  
and spatial resolution capability. In the present letter  we study 
theoretically one possible way to enhance the optical transmittance 
through a metallized near-field probe with a subwavelength aperture. 
We suggest  to work in the visible region using a short probe with a large taper angle with a core consisting 
of a semiconducting matter with a high refractive index $n$.  The increase in transmittance with $n$ 
occurs due to decrease of the light wavelength $\lambda _{c} = 2\pi c/n\omega$ inside 
the core. 
Consequently, at high $n$ 
the cutoff effect, which strongly reduces the light transmission through the  
probe in the overdamped regime, affects its transmission efficiency much 
less than for a glass. However, this gain  may be counterbalanced by the growth of  light absorption in the dissipative 
medium of the core described by the imaginary part of the 
frequency dependent dielectric function $\varepsilon(\omega)$. To deal 
with these two competing tendencies one needs a detailed theory.  

The purpose of this letter is to  develop an  
analytic theory for transmission of visible light through a probe with a 
core made of a dissipative semiconducting matter. Here we are encouraged by  the transmission  SNOM experiments with a silicon probe  at $\lambda = 1.06$ $\mu$m
\cite{Danzebrink98, Dziomba2001}, which  indicate that such probes are very 
promising in the near-infrared (IR) region. The theoretical 
support for this work comes from  a comparative numerical analysis of the transmission efficiencies 
of glass and silicon probes at $\lambda=1.3$ $\mu$m 
\cite{Castiaux98}. We have to note however, that the above theory  
 is restricted by the use of a two-dimensional model with 
a loss-free dielectric core and small taper angle of $15^{\circ}$.  
The further experimental success  
was achieved in \cite{Yatsui2002}, where the authors 
have employed a pyramidal Si probe that was entirely coated with a 
thin metal film to increase the transmission efficiency in the near-IR   
region ($\lambda =830$ nm). An extremely high throughput ($2.3$ \%) 
was obtained  in this experiment with a  resolution capability about 
$85$ nm. In the IR  region the light absorption in Si is 
sufficiently small, but  in the visible region the imaginary part of its 
dielectric function rapidly increases with a decrease of $\lambda$. So, 
 dissipation of the electromagnetic energy inside the silicon core 
and frequency dispersion of its dielectric function 
become important 
and should be taken into account to get an adequate physical pattern of 
light transmission through the probe. 

 The three-dimensional theory developed in the present 
letter is valid for all taper angles including large ones which are especially 
suitable for large transmittance \cite{Novotny95, KLT03}. The theory is 
based on the exact analytic description of the conical waveguide eigenmodes 
inside a probe with a dissipative matter in its core and perfectly conducting 
metallic walls. For a loss-free dielectric core, similar approach  has been 
developed in our recent work \cite{KLT03}. Here  we consider  
the case when the semiconducting core has a complex dielectric function 
which is a necessary feature of the visible region. This generalization  
requires more than a trivial analytic continuation of our previous results 
due to the necessity of describing the effects associated with frequency 
dispersion and light absorption inside the core matter. Such effects 
lead to a number of new  features in light transmission through 
semiconducting probes, which discuss below.

 We consider here time-harmonic fields inside a cone whose core 
consists of a dissipative medium and whose walls are perfectly conducting. 
In spherical coordinates the Helmholtz equation for the Hertz function $U$ 
of the electromagnetic field inside a cone is  

\begin{equation}
\frac{\partial ^{2}U}{\partial r^{2}}+\frac{1}{r^{2}}\left[ \frac{1}{\sin
\theta }\frac{\partial }{\partial \theta }\left( \sin \theta \frac{\partial U%
}{\partial \theta }\right) +\frac{1}{\sin ^{2}\theta }\frac{\partial ^{2}U}{%
\partial \varphi ^{2}}\right] +k^{2}U=0.  \label{w-eq}
\end{equation}

\noindent Here $r$ is the distance from the cone vertex, $\theta $ and 
$\varphi $ are the polar and azimuthal angles. 
For a dissipative  matter, the wave number $k$ is complex: $
k=\omega \sqrt{\varepsilon \mu }/c$, $\sqrt{\varepsilon \mu }=n+i\kappa$,  
where  $c$ is the speed  of light, $n$ and $\kappa $ are the refractive 
index and the attenuation coefficient. Assuming the 
permeability $\mu =1$, the real 
$\varepsilon'=\Re e\left\{ \varepsilon \right\} $ and imaginary 
$\varepsilon''=\Im m\left\{ \varepsilon \right\} $ parts of the 
frequency-dependent dielectric function 
$\varepsilon \left( \omega \right) =\varepsilon'\left( \omega
\right) +i\varepsilon''\left( \omega \right)$ 
are given by $\varepsilon'=n^{2}-\kappa ^{2}$ and 
$\varepsilon''=2n\kappa $ (see Ref.\cite{LL}).

The relevant solution of Eq. (\ref{w-eq}) corresponding  to the standing wave 
with vanishing amplitude at the cone vertex ($r=0$), has the form

\begin{equation}
U=\mathcal{R}\left( r\right) P_{\nu }^{m}\left( \cos
\theta \right) e^{im\varphi }\,, \quad
\mathcal{R}=Crj_{\nu }\left(kr\right) \,,
\label{potent}
\end{equation}

\noindent at which the radial dependence $\mathcal{R}\left( r\right) $ of 
the Hertz function (\ref{potent}) is expressed through the spherical Bessel 
function of the first kind $j_{\nu }\left( z\right)$ of a complex argument 
with the index $\nu $ not equal to an integer. Here $C$ is a constant. 
The dependence on the polar angle $\theta $ is determined by the associated 
Legendre function of the first kind 
$P_{\nu }^{m}\left( \cos \theta \right) $ with power $\nu $ and order $m$ 
($m$ is an integer). 

For the transverse magnetic (TM) field modes 
the boundary condition at an interface between a core medium and a 
perfectly conducting metallic coating of a conical waveguide can be written 
as $P_{\nu }^{m}\left( \cos \theta _{0}\right) =0$, 
where $\theta _{0}$ is the cone half-angle. Each choice of numbers $m$ 
and $n$ in this equation ($n$ denotes the number of its root) 
determines a possible TM$_{mn}$ mode.  
The eigenvalues $\nu _{mn}$ strongly depend upon the value of 
$\theta _{0}$ such that $\nu_{mn}$ decreases with an increase of $\theta_0$. 
In the most interesting case of the lowest-order TM$_{01}$ mode, the 
projections of electric $\mathbf{E}$ and magnetic $\mathbf{H}$ fields onto 
the corresponding axes of spherical coordinates $(r,\,\theta ,\,\varphi )$, 
take the form  

\begin{equation}
E_{r}=\frac{\nu \left( \nu +1\right) }{r^{2}}\,\mathcal{R}(r)P_{\nu }\left(
\cos \theta \right), ~~
E_{\theta }=\frac{\partial \mathcal{R}(r)}{r\partial r}\frac{%
\partial P_{\nu }\left( \cos \theta \right) }{\partial \theta },
\label{E_theta}
\end{equation}

\begin{equation}
H_{\varphi }=i\frac{\omega \left( \varepsilon ^{\prime }+i\varepsilon
^{\prime \prime }\right) }{c}\frac{1}{r}\mathcal{R}(r)\frac{\partial P_{\nu
}\left( \cos \theta \right) }{\partial \theta }\,.\label{H_phi}
\end{equation}

\noindent while the $H_{r}$, $H_{\theta}$, and $E_{\varphi}$ components 
are equal to zero.

In a dissipative media with frequency-dependent dielectric function 
$\varepsilon =\varepsilon'+\varepsilon'' $ 
and permeability $\mu =\mu '+i\mu ''$, the general expressions (see 
\cite{LL}) for the time-averaged densities of the electric,  
$w_{el}=w_{r}+w_{\theta }$, and magnetic, $w_{m}=w_{\varphi }$, fields  
are given by

\begin{equation}
w_{el}=\frac{1}{16\pi }\frac{d\left( \omega \varepsilon '\right) }{d\omega }
\left( \left| E_{r}\right| ^{2}+\left| E_{\theta}\right| ^{2} \right)\,,  
\label{w_el}
\end{equation}

\begin{equation}
w_{m }=\frac{1}{16\pi }\frac{d\left( \omega \mu '\right) }
{d\omega }\left| H_{\varphi }\right| ^{2},  \label{w_phi}
\end{equation}
  
\noindent To determine the near-field transmission coefficient of a 
truncated conical waveguide we introduce the quantities 

\begin{equation}
W_{\beta}(r)=2\pi r^{2}\int\limits_{0}^{\theta _{0}}
w_{\beta}\left( r,\theta \right)
\sin \theta d\theta \,,   
\label{W_beta}
\end{equation}

\noindent which represent the integrals of 
$w_{r}$, $w_{\theta }$, or $w_{\varphi }$ 
taken over a  part of spherical surface lying inside the  cone 
($0\leq \theta \leq \theta _{0},\;0\leq \varphi \leq 2\pi $)\ at a given 
distance $r$ from the cone vertex. With the help of Eqs. 
(\ref{potent}), and (\ref{E_theta})--(\ref{W_beta}), these integrals can be 
evaluated explicitly. 
The resulting expressions for $W_r$,  $W_{\theta}$, and $W_{\varphi}$, 
take the form

\begin{eqnarray}
&& W_{r}=\frac{\left| C\right| ^{2}}{8}\frac{d\left( \omega
\varepsilon ^{\prime }\right) }{d\omega } \nu \left( \nu +1\right)
\mathcal{I}_{\nu }\left| j_{\nu }\left[
\left( n+i\kappa \right) \frac{\omega r}{c}\right] \right| ^{2},
\label{W_r}
\end{eqnarray}
\begin{eqnarray}
 &&W_{\theta }=\frac{\left| C\right| ^{2}}{8}\frac{d\left(
\omega \varepsilon ^{\prime }\right) }{d\omega }
\mathcal{I}_{\nu } 
\left| 
\left( \nu +1\right) j_{\nu }\left[ \left( n+i\kappa
\right) \frac{\omega r}{c}\right]\right . \nonumber\\
&& \left .  -\left[ \left( n+i\kappa \right) \frac{%
\omega r}{c}\right] j_{\nu +1}\left[ \left( n+i\kappa \right) \frac{%
\omega r}{c}\right]\right| ^{2},
\label{W_theta}
\end{eqnarray}
\begin{eqnarray}
&& W_{\varphi } =\frac{\left| C\right| ^{2}\left| \varepsilon \right| }
{8}\left( \frac{\omega r}{c}\right) ^{2}
\mathcal{I}_{\nu }
\left| j_{\nu }\left[ \left( n+i\kappa \right) \frac{\omega r}{c}\right]
\right| ^{2}.  \label{W_m}
\end{eqnarray}

\noindent Here the angular integral $\mathcal{I}_{\nu }$ 
is given by 

\begin{equation}
\mathcal{I}_{\nu }=\int\limits_{0}^{\theta _{0}}\left[ 
\frac{\partial P_{\nu }\left( \cos \theta \right) }{\partial \theta }\right]
^{2}\sin \theta d\theta \,.  
\label{ang_int2}
\end{equation}

The integral energy density $W_{tot}=W_{r}+W_{\theta}+W_{\varphi}$  
can now be evaluated as the sum of Eqs. (\ref{W_r})--(\ref{W_m}). At small 
distances from 
the cone vertex ($r \ll \lambda_c$) it  
exhibits a rapid power fall, $W_{tot} \propto (|k|r)^{2\nu}$, with a 
decrease of $r$. At distances $r$ from the cone vertex much 
greater than the wavelength in the core medium ($r\gg \lambda_{c}$), 
the asymptotic expression for $W_{tot}$ takes the form 
 
\begin{equation}
\begin{array}{l}
W_{tot}=
\frac{\left| C\right| ^{2}}{16}
\mathcal{I}_{\nu }\left\{ \left[ 
\frac{d\left( \omega \varepsilon ^{\prime }\right) }{d\omega }+\left|
\varepsilon \right| \right] 
\cosh \left( \frac{r}{r_{\kappa}}\right)\right. \\ 
\\ 
+\left. \left[ \frac{d\left( \omega \varepsilon ^{\prime }\right) }{d\omega }%
-\left| \varepsilon \right| \right] \cos \left( 2n\frac{\omega r}{c}-\pi \nu
\right) \right\} .
\end{array}
\label{W_tot_as}
\end{equation}

\noindent Here $r_{\kappa}=c/2\kappa \omega $ is the attenuation length. 
It is evident that the main feature in the radial dependence 
(\ref{W_tot_as}) is determined by a factor 
$\cosh\left(r/r_{\kappa}\right)$, which reflects the influence of 
light absorption inside a dissipative core of a near-field probe. 
Another important point is that  the integral energy density 
$W_{tot}(r)$ exhibits an oscillatory behavior at distances far from the 
cone vertex $r\gg c/\omega \left| n+i\kappa \right|$. These additional 
oscillations are the result of the frequency-dependent dielectric 
function; they are absent if the core is made of a loss-free medium 
(this is the case of a probe with a glass core, 
$\varepsilon =const$ and $\kappa =0$, see \cite{KLT03}). 
However, for a lossy matter the amplitudes of the electric 
($W_{el}\propto d\left( \omega \varepsilon' \right)/d\omega $,  
Eqs. (\ref{W_r}), (\ref{W_theta})) and magnetic 
($W_{m}\propto \left| \varepsilon \right|$, 
Eq. (\ref{W_m})) components of the integral energy densities are not equal and 
can be significantly differed from each other. Therefore, the oscillations 
of the electric, $W_{el}$, and magnetic, $W_{m}$, energies do not 
compensate each other far from the cone vertex. Note also that in the 
presence of light absorption there is a significant 
difference in magnitudes of the energy fluxes associated with the 
incident and the reflected waves. Their ratio $\left|S_r/S_{in}\right|$ 
is equal to $\exp\left(-2r/r_{\kappa}\right)$. 
So, this difference becomes particularly important in the range of 
$r\gg r_{\kappa}$.

Further we evaluate the optical transmittance of a conical waveguide 
with a dissipative matter in its core. For a probe tapered to a 
subwavelength diameter it is necessary to distinguish a near-field 
transmission coefficient of a waveguide itself and the resulting transmission 
coefficient to the far-field zone (see \cite{Hecht2000, KLT03}).     
The near-field transmission coefficient, $T$, can be expressed in terms of 
the time-averaged energy densities associated with the output and the input 
fields of the waveguide. For  spherical waves inside a cone, 
this coefficient 
can be defined as the ratio $T=W_{tot}^{out}/W_{tot}^{in} $
of the time-averaged energy density $W_{tot}^{out}\equiv 
W_{tot}\left( z_{0}\right) $ at the exit plane $z=z_{0}$ of a
probe integrated over the aperture cross section $2\pi \rho d\rho $ 
with radius $a$ ($a=z_{0}\tan \theta _{0}$) to the corresponding 
integral energy density $W_{tot}^{in}$ at the waveguide entrance with 
radial coordinate $r_{in}$. In a dissipative medium, the latter is given 
by $W_{tot}^{in}=\alpha W_{tot}\left(r_{in} \right)$ (see (\ref{W_beta})). 
The factor 
$\alpha =\left[1+\exp \left( -2r_{in}/r_{\kappa}\right)\right]^{-1}$  
shows a fraction of the integral energy density at $r=r_{in}$,  
associated with the incident wave alone. So, the contribution 
of the reflected wave turns out to be completely removed. 

For the subwavelength aperture $2a\ll \lambda _c$,  
the basic expressions (\ref{W_r})--(\ref{W_m}) for $W_{r}$, $W_{\theta}$, 
and $W_{\varphi}$ 
at the exit of a conical waveguide, can be 
expanded in power series of $\left|k\right|r_{out}$, where $r_{out}=a/\sin\theta_0$ is the corresponding radial coordinate. For the value of 
$W_{tot}\left(r_{in}\right)$ at the waveguide entrance ($r_{in}\gg \lambda_c$) 
we use the expression 
$\overline{W}_{tot}\left(r\right) \propto \cosh \left( r/r_{\kappa}\right)$,  
averaged over the fast oscillations of Eq. (\ref{W_tot_as}). Then, the 
resulting expression for the near-field transmission coefficient 
takes the form

\begin{equation}
T\propto 
\left( \frac{\omega \left|n+i\kappa \right| a}
{c \sin \theta_0}\right)^{2\nu\left(\theta_0\right)}
 \cosh^{-1} \left[\frac{ r_{in}}{r_{\kappa}\left(\omega\right)}\right]\, .
\label{T_anal}
\end{equation}

\noindent The eigenvalues $\nu\equiv \nu_{01}$ of the TM$_{01}$ 
mode in Eq. (\ref{T_anal}) exhibit rapid fall with an increase of the 
taper angle ($\nu_{01}=4.083$,  $2.548$, $1.777$, and $1$ at 
$\theta_0=\pi/6$,  $\pi/4$, $\pi/3$, and $\pi/2$, respectively). 
Thus, Eq. (\ref{T_anal}) describes well all major features of 
light transmission through the subwavelength aperture in a conical 
waveguide with a dissipative matter in its core. It is seen 
that the values of $T$ are strongly dependent on the ratio $a/\lambda$, 
the taper angle $2\theta_0$, and the refractive index $n$. 
Moreover, according to (\ref{T_anal}) the transmission coefficient 
$T$ is proportional to $\cosh^{-1} \xi $, where $\xi=l/r_{\kappa}$ 
is the ratio of the length of the probe edge $l$ to the  
attenuation length $r_{\kappa}=c/2\kappa \omega$ 
(at small $a$ we have $r_{in} \approx l$). It is clear that  the high 
transmission efficiency of a semiconducting probe can be achieved in the 
wavelength region far from the peak in its absorption band ($\kappa\ll n$). 
If additionally $r_{\kappa} \gg l$, than one can put 
$r_{\kappa}\rightarrow \infty $. Then, Eq. (\ref{T_anal}) is reduced 
to especially simple form $T \propto 
\left(\omega n a/c \sin\theta_0\right)^{2\nu\left(\theta_0\right)}$. 
This is the case of a loss-free dielectric core. In the opposite case 
of large losses ($r_{\kappa} \ll l$), the transmission coefficient   
behaves like $T\propto \left(\omega n a/c \sin\theta_0\right)^{2\nu\left
(\theta_0\right)}\exp \left(-\xi\right)$. This reflects the 
strong influence of light absorption on the value of $T$.

Now we apply our theory for studies of the 
transmittance of the visible and near-IR radiation through the 
aperture-type metallized silicon probe. In the wavelength region from 830 nm ($\hbar\omega = 1.5$ ev)
down to 400 nm ($\hbar\omega = 3.1$ ev), the refractive index $n$ of Si increases monotonically 
from $3.67$ to $5.57$ and the attenuation coefficient $\kappa $ grows 
from $0.005$ to $0.387$.  
This results in quite different influence of light absorption in Si 
in the near-IR and the short-wavelength part of the visible 
spectrum. 

In Fig. 1 we present the wavelength dependences of the near-field transmission 
coefficient of the metallized silicon probe for the most interesting case of 
large taper angle $2\theta_0=90^{\circ}$ and for various values of the 
aperture diameter $2a$. To demonstrate a dependence  of light absorption 
inside the Si core on the length of the probe edge, 
 we calculated the transmission coefficient $T$  for various values of $l$. As expected, $T$ is strongly dependent on the aperture diameter 
in full agreement with simple formula (\ref{T_anal}) derived in this work. 
However, the wavelength dependence, obtained in the present work 
for the Si probe, differs dramatically from the case of a loss-free dielectric 
core (a glass fiber), for which $T\propto \left(a/\lambda\right)^{2\nu}$). 
As is evident from Fig. 1, the transmittance of the silicon probe strongly 
varies over the spectrum. If the length of the probe edge is not too large 
($l\lesssim 10$ $\mu$m), the transmission coefficient 
increases first as the wavelength decreases from the IR region, 
reaches its maximum at a definite wavelength $\lambda_{\max}$, and then 
strongly falls at $\lambda \ll \lambda_{\max}$ in the short-wavelength part 
of the visible spectrum. The position of the maximum $\lambda_{\max}$ and the 
maximal value of $T_{\max}$ depends on the specific geometrical parameters of 
the probe. 

\begin{figure}[ht]
\begin{center}
\epsfxsize=0.4\textwidth
\epsfbox{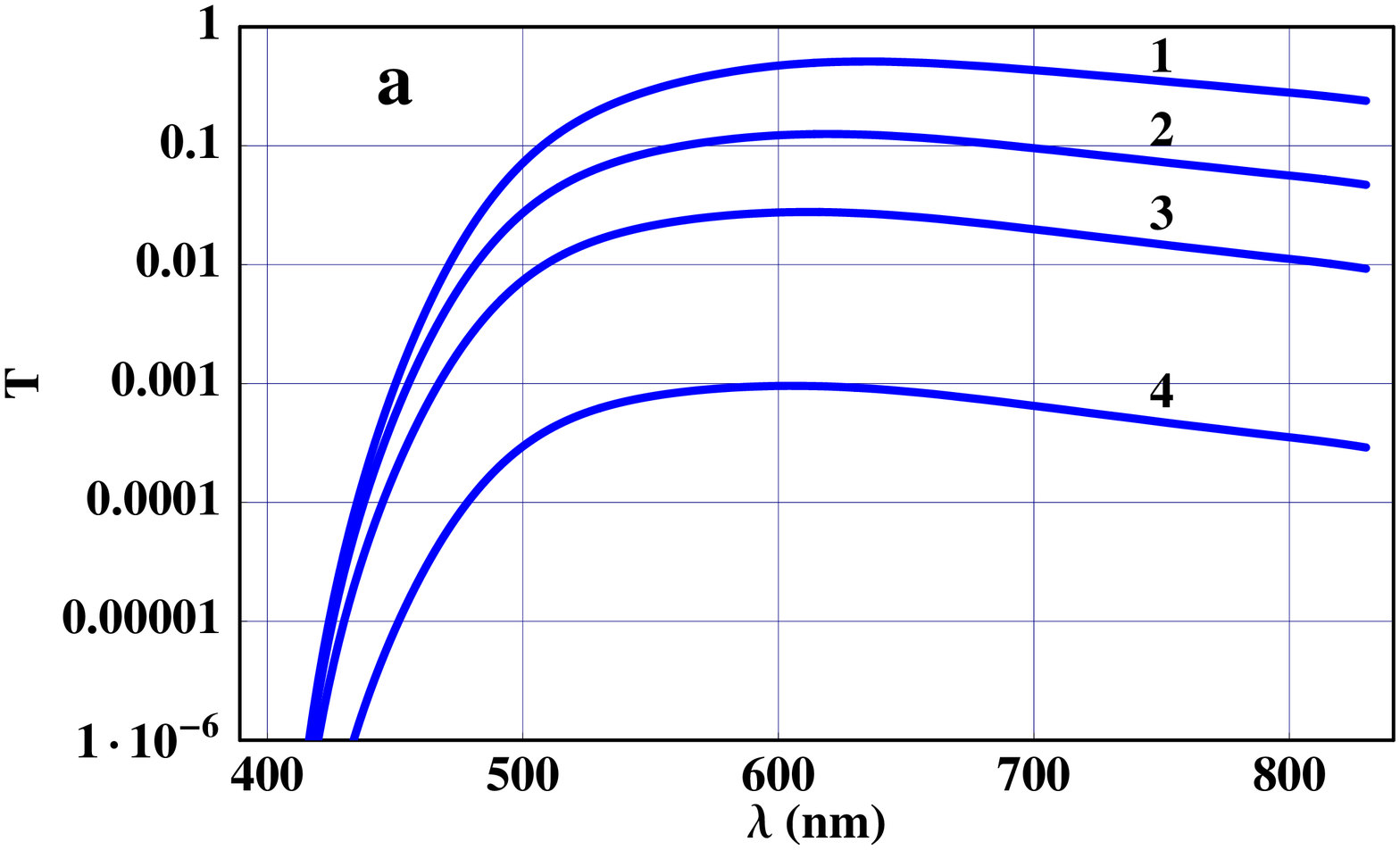}
\epsfxsize=0.4\textwidth
\epsfbox{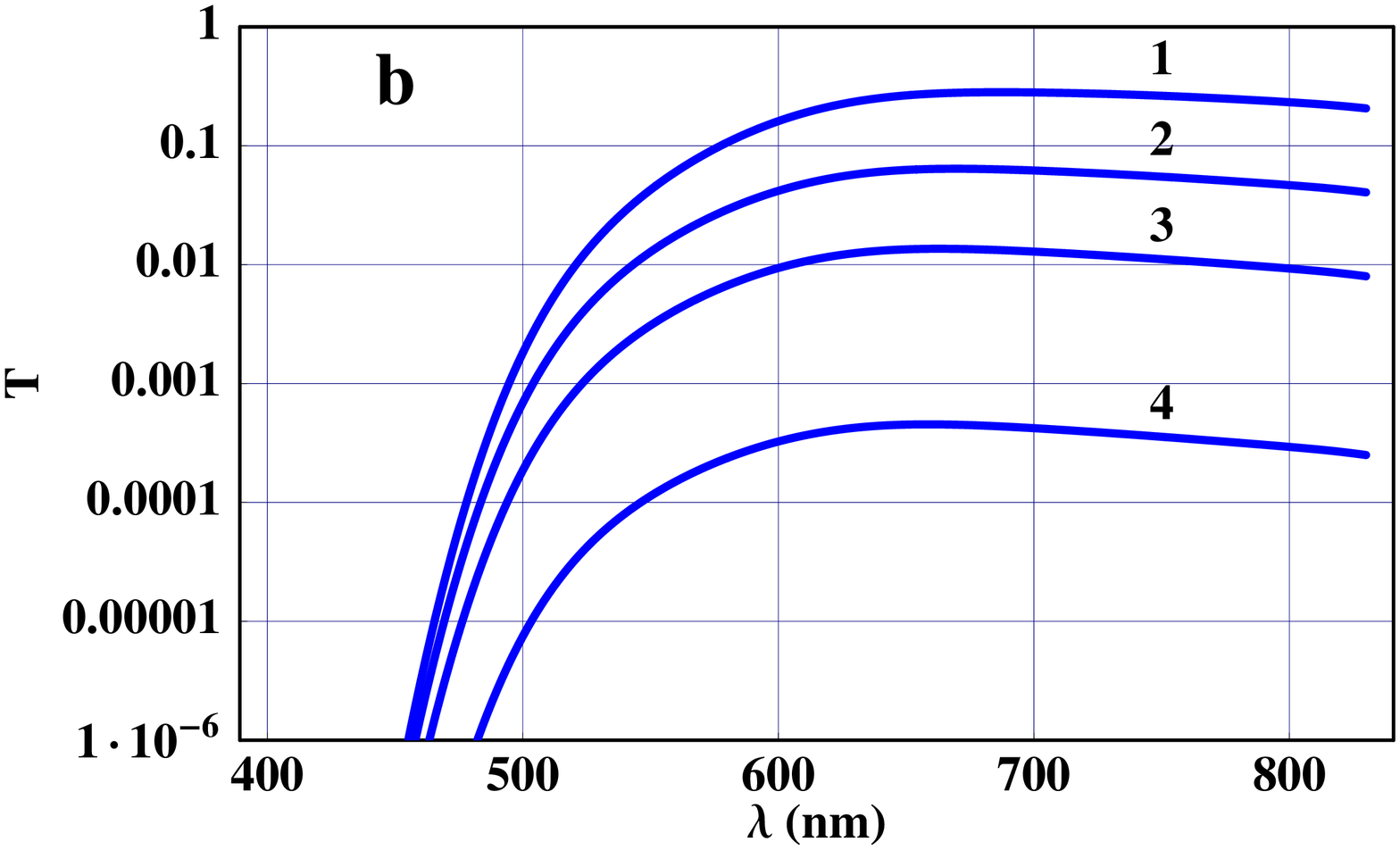}
\epsfxsize=0.4\textwidth
\epsfbox{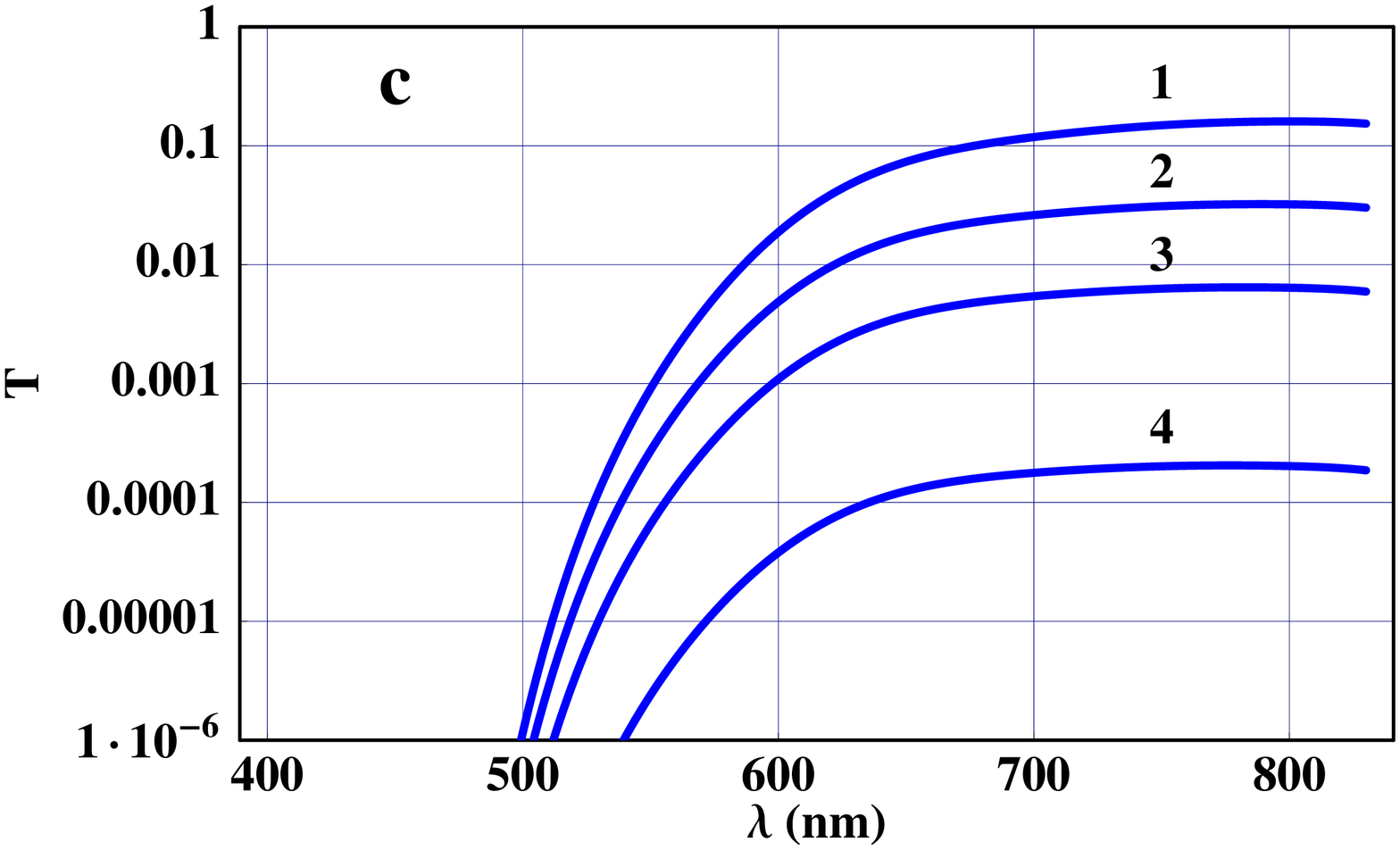}
\end{center}
\caption{The near-field transmission coefficient, 
$T$, of the conical waveguide  ($2\theta _{0}=90^{\circ}$) 
with the Si core {\it vs}  $\lambda =2\pi c/\omega$. Curves 1, 2, 3, and 4 
correspond to the aperture diameter $2a=100$ nm, $70 $ nm, $50$ nm, 
and $25$ nm, respectively. The  
length of the probe edge is $l=2$ $\mu$m (a), 4 $\mu$m (b), 8 $\mu$m (c).} 
\end{figure}

It is important to stress that this maximum in the transmission efficiency 
of a silicon probe lies at $\lambda \sim 550-800$ nm (see Fig. 1). This occurs 
despite the fact that the attenuation 
length $r_{\kappa}$, associated with the imaginary part of the dielectric 
function of Si, considerably decreases in the visible region compared to the 
near-infrared one. For example, at $\lambda =633$ nm,  
$532$ nm, and $488$ nm, the respective values of the attenuation length 
$r_{\kappa}$ turn out to be equal to $2.66$ $\mu$m, 0.84 $\mu$m 
and 0.45 $\mu$m, in contrast with  
13.53 $\mu$m at $\lambda=830$ nm.

In summary, it follows from our calculations that at large taper 
angles high values of the near-field transmission coefficient 
can be achieved for a  passage of visible light through Si core 
of an optical probe. To illustrate the enhancement in the trasmittance of Si 
probes in comparison with conventional fiber ones we compare the present 
estimates for Si with those obtained for a core with small 
$n$ (glass or SiO$_2$). 
Although the taper angles of fiber probes do not usually exceed 
40$^{\circ}$ which  additionally restricts their efficiency, we use 
the value $2\theta_0=90^{\circ}$ and $n = 1.55$ to 
make comparison with our recent results \cite{KLT03}.       
For the probe with the length $l=2$ $\mu$m and the aperture diameter 
$2a =50$ nm we get 
$T_{Si}/T_{glass} = 2.2$, 14, 45, and 71 for $\lambda = 488$, 532, 633, 
and 830 nm, respectively. For the same parameters, but $2\theta_0=60^{\circ}$ 
we have $T_{Si}/T_{glass} =43$, 240, 800, and 960. According to our theory, 
the enhancement  occurs as a result of competition between two factors: the 
rise of $n$ and the decrease of the attenuation length $r_{\kappa}$ (\ref{T_anal}). As follows from our results, the former factor is more 
important in the most part of the visible spectrum, provided the near-field 
probe length is sufficiently short (no more than several $\mu$m), such that 
effects associated with the light absorption are not too strong. 
We also would like to point out that in case of an entirely coated 
probe the surface plasmon-polariton propagation along the metallic 
cladding is likely to further increase the resulting transmittance in 
accordance with the  mechanism discussed in Ref. \cite{Novotny95}.

This work was supported in part by the Programme ''Optical Spectroscopy 
and Frequency Standards'' of the Division of Physical Sciences of the Russian 
Academy of Sciences.   
AMT acknowledges the support from US DOE under contract number 
DE-AC02 -98 CH 10886. VSL acknowledges a support 
from Institute for Strongly Correlated and Complex Systems at BNL.

\end{document}